\documentclass[aps,prb,twocolumn,amsmath,amssymb,superscriptaddress,floatfix]{revtex4}
\usepackage{graphicx}
\usepackage{bm}
\usepackage[usenames]{color}
\usepackage{colordvi}
\bibstyle{apsrev.bib}

\newcommand{\be}{\begin{equation}}
\newcommand{\ee}{\end{equation}}
\newcommand{\bea}{\begin{eqnarray}}
\newcommand{\eea}{\end{eqnarray}}
\newcommand{\beqn}{\begin{eqnarray}}
\newcommand{\eeqn}{\end{eqnarray}}

\begin{document}

\title{Semi-classical theory for quantum quenches in
finite transverse Ising chains}
\author{Heiko Rieger}
\email{h.rieger@mx.uni-saarland.de}
\affiliation{Theoretische Physik, Universit\"at des Saarlandes, 66041 
Saarbr\"ucken, Germany}
\author{Ferenc Igl\'oi}
\email{igloi@szfki.hu}
\affiliation{Research Institute for Solid State Physics and Optics,
H-1525 Budapest, P.O.Box 49, Hungary}
  \affiliation{Institute of Theoretical Physics,
Szeged University, H-6720 Szeged, Hungary}
\date{\today}

\begin{abstract}
We present a quantitative semi-classical theory for the
non-equilibrium dynamics of transverse Ising chains after quantum
quenches, in particular sudden changes of the transverse field
strength.  We obtain accurate predictions for the quench dependent
relaxation times and correlation lengths, and also about the
recurrence times and quasi-periodicity of time dependent correlations
in finite systems with open or periodic boundary conditions.  We
compare the quantitative predictions of our semi-classical theory (local
magnetization, equal time bulk-bulk and surface-to-bulk correlations,
and bulk autocorrelations) with the results from exact free fermion
calculations and discuss the range of applicability of the
semi-classical theory and possible generalizations and extensions.
\end{abstract}


\maketitle

\section{Introduction}

The non-equilibrium quantum relaxation in many-body systems has gained
increased interest over the recent years, not least because trapped
cold atom systems made its experimental study possible.  In principle
one asks for the fate of an initial state that is not an eigenstate of
the Hamiltonian under the time evolution according to the
Schr\"odinger equation.  A straightforward method to prepare such an
initial state is the instantaneous change of a global or local
parameter of the system like an external field or the interaction
strength, denoted as a quantum quench or simply quench.  Important
issues of interest are then: 1) Is there an asymptotic stationary
state, what are its characteristics, is it describable by a general
Gibbs ensemble (i.e.\ does the system thermalize after a quench)?
2) What are the characteristics of the dynamical evolution of order,
correlations and quantum entanglement in the system?

The first theoretical studies of quenches in quantum many body systems
were performed for the quantum XY and quantum Ising spin chains
\cite{barouch_mccoy,igloi_rieger,sengupta}. Spectacular experimental
results \cite{exp} triggered an intensive research on quantum quenches
in various systems such as 1D Bose systems \cite{rigol}, the quantum
sine-Gordon model \cite{gritsev}, Luttinger liquids \cite{cazalilla}
and others \cite{manmana}. Besides studies on specific models there
are also field-theoretical investigations, in which relation with
boundary critical phenomena and conformal field-theory are utilized
\cite{calabrese_cardy,sotiriadis_cardy,gambassi}.  Progress in
understanding thermalization, or absence thereof, in a particularly
well studied integrable model, the transverse Ising chain, has been
achieved in \cite{rossini,igloi_rieger2}. The concept of an effective
temperature depending on the quench parameters is useful to
parametrize the relaxation time and correlation length determining the
spin correlations after a global quench. But actually each excitation
mode has its own thermalization temperature \cite{Calabrese2},
implying that the system never thermalizes after a quench.

For the transverse Ising chain in thermal equilibrium Sachdev and Young
\cite{sachdev_young} introduced a semi-classical description of the
equilibrium quantum relaxation in terms of ballistically moving
quasi-particles. This description turned out to be surprisingly
accurate in predicting the temperature dependence of relaxation time,
correlation length and scaling forms in the ferromagnetic and
paramagnetic phase. 

For global quantum quenches a picture of ballistically moving quasi-particles
spontaneously created after the
quench has been used\cite{calabrese_cardy,calabrese_cardy2} to explain several
features of the time-evolution of different quantities, in particular that
of the entanglement entropy\cite{fagotti_calabrese,gradient}. This picture has also been used to interpret results of exact
calculations obtained with the free fermion technique
\cite{rossini,igloi_rieger2} or field theory (at the critical point)
\cite{calabrese_cardy,calabrese_cardy2}. 

Obviously it would be desirable to have a
quantitative semi-classical theory for the non-equilibrium dynamics
after quantum quenches, too.
This is what we will present in this paper
for global quenches, for local quenches a brief account has been
given by us recently in \cite{uma}.
Here we present the quantitative
analog of the semi-classical theory for equilibrium quantum
relaxation of transverse Ising chains
\cite{sachdev_young} and generalize it to the non-equilibrium 
dynamics in {\it finite} systems. By this we will not only obtain
accurate predictions for the relaxation times and correlation lengths,
but also about the recurrence times and quasi-periodicity of time
dependent correlations in finite systems with open or periodic
boundary conditions.
Since in experimental set-ups of quantum
quenches, as for instance cold atom systems, the number of
particles is rather restricted and far away from the
infinite system size limit, the understanding of 
finite size effects in non-equilibrium quantum
relaxation is important and may be, as we will show 
often be drastic.

The paper is organized as follows: After the model definition in the
next section we present the semi-classical theory for the
non-equilibrium dynamics of the transverse Ising chain after a quench.
Then we derive the semi-classical formula for the local
magnetization, equal time bulk-bulk and surface-to-bulk correlations,
and bulk autocorrelations and compare the predictions with the
results from exact free fermion calculations. Finally we discuss
the range of applicability of the semi-classical theory and
possible generalizations and extensions.

\section{Model}
\label{sec:model}

The system we consider in this paper is the quantum Ising chain
defined by the Hamiltonian\cite{pfeuty}:
\be
{\cal H}=-\dfrac{1}{2}\sum_{l=1}^{L-1} \sigma_l^x \sigma_{l+1}^x -
\dfrac{h}{2} \sum_{l=1}^{L} \sigma_l^z\;,
\label{hamilton}
\ee
in terms of the Pauli-matrices $\sigma_l^{x,z}$ at site $l$. In
(\ref{hamilton}) the chain has a finite length $L$ and open
boundaries, later we will also discuss periodic boundary
conditions. We consider global quenches in which the transverse field
strength is suddenly changed from $h_0$ for $t<0$ to $h\ne h_0$ for
$t>0$.  For $t<0$ the system is in equilibrium, which means it is in
its ground state $|\varPsi_0\rangle$ and which we denote as its
initial state. After the quench, for $t>0$, the state
$|\varPsi_0\rangle$ evolves according to the new Hamiltonian:
\be
|\varPsi_0(t)\rangle = \exp\left(-\imath{\cal H}t\right) |\varPsi_0\rangle\;.
\label{Psi}
\ee
Similarly we have for the time-evolution of an operator:
$\sigma_{l}(t)=\exp(-\imath t {\cal H}) \sigma_{l} \exp(\imath t {\cal H})$.

We consider the general, time and space dependent correlation function:
\be
C(r_1,t_1;r_2,t_2)=
\langle\varPsi_0|\sigma_{r_1}^x(t_1)\sigma_{r_2}^x(t_2)
|\varPsi_0\rangle\;,
\label{C}
\ee
and study its behavior in special circumstances. The autocorrelation
function is obtained for $r_1=r_2=r$, which is denoted as
$G_r(t_1,t_2)$, whereas for $t_1=t_2=t$ we have the equal-time
correlation function. This latter quantity for large separation
behaves as: $C(r_1,t;r_2,t)\equiv C_t(r_1,r_2)=m_{r_1}(t)m_{r_2}(t)$,
where $m_{r}(t)$ is the local magnetization. In the initial state (and
in the thermodynamic limit, $L \to \infty$) for $h_0<h_c=1$ there is a
finite magnetization, $m_r(0)>0$, whereas for $h_0>1$ one has 
$m_r(0)\sim{\cal O}(1/L)$.

\subsection{Free fermion representation}
\label{sec:free_ferm}

The Hamiltonian in Eq.(\ref{hamilton}) can be expressed in terms of
free fermion creation, $\eta_p^{\dag}$, and annihilation operators,
$\eta_p$\cite{lieb,pfeuty} :
\be
{\cal H}=\sum_{p} \varepsilon_h(p)\left(\eta_p^{\dag}\eta_p-1/2\right)\;,
\label{hamilton_free}
\ee
where the energy of modes is given by
\be
\varepsilon_h(p)=\sqrt{(h-\cos p)^2+sin^2 p}\;.
\label{epsilon}
\ee
The quasi-momenta, $p$, has $L$ quasi-equidistant values in the
interval: $0<p<\pi$ for free boundary conditions, whereas for closed
chains these are restricted to $|p|<\pi$.  Time-evolution of the
fermion operators are $\eta_p^{\dag}(t)=e^{\imath t\epsilon_k}
\eta_k^{\dag}$ and $\eta_k(t)=e^{-\imath t\epsilon_k} \eta_k$ from
which one can obtain the time-evolution of the spin operators. The
correlation functions in the fermion representations are expressed
in terms of Pfaffians, which are then calculated as the square-root of
the determinant of the corresponding antisymmetric matrix, which has
the elements of the Pfaffian above the diagonal. For free boundary
conditions these determinants have a dimension:
$2(r_1+r_2)$. Following Yang\cite{yang}, the local magnetizations can be
calculated in the form of an off-diagonal matrix-element:
$m_r(t)=\langle \varPsi_0 | \sigma_r^x(t) | \varPsi_1 \rangle$, where
$|\varPsi_1\rangle$ denotes the first excited state for $t<0$. Its numerical
calculation necessities the solution of a $2r \times 2r$ determinant.

\section{Semi-classical theory}
\label{sec:QP}

In the absence of the transverse field in Eq.(\ref{hamilton}), $h=0$,
the system is, in the transformed basis $\sigma_i^x \leftrightarrow
\sigma_i^z$ identical with the classical Ising spin chain.  The ground
state is two-fold degenerate and given by:
$|\varPsi_0\rangle=|+++\dots+\rangle$ and
$|\varPsi_0\rangle=|---\dots-\rangle$ and the first excited states are
the $(L-1)$-fold degenerate given by the single kink states
$|n\rangle=|++\dots++--\dots--\rangle$, where $n$ denotes the kink
position. Switching on a small transverse field, $h>0$, the low-lying
excitations are, in first order order degenerate perturbation theory,
superpositions of these single-kink states $\sum_n a_n |n\rangle$ with
excitation energy $\varepsilon_h(p)$.
%They can be calculated by comparing the total energy of
%the system with parallel ($++$) and anti-parallel ($+-$) fixed spin
%boundary conditions for open chains. (Similarly, one can make the
%calculation with periodic and anti-periodic boundary conditions for
%closed chains).
The actual perturbation calculation yields $a_n=\sqrt{2/L}\sin(pn)$,
with $\varepsilon_h(p)=1-h\cos p$, where $p$ has $L-1$ discrete values
in the same region as given below Eq.(\ref{epsilon}).  Thus the low
lying excitations of ${\cal H}$ are Fourier transforms of localized
single kink states, similar to the eigenstates of the Hamiltonian for
free particles in a box of length $L$. Analogously freely moving
single kinks are therefore wave packets of the aforementioned low
lying excitations.  Their energy agrees to to leading order in $h$
with the free-fermion energies in Eq.(\ref{epsilon}) and they move
ballistically with constant velocity $\pm v_p$ given by
\be
v_p=\frac{\partial\varepsilon_p}{\partial p}
=\frac{h\sin(p)}{\epsilon_p}\;.
\label{vp}
\ee

Ballistically moving kinks are then the (fermionic) quasi-particles
(QPs) which we use in the following to formulate a semi-classical
theory of the quantum quench dynamics of the transverse Ising
model. Since by definition these QPs are well-defined at small fields
in the ferromagnetic phase, the theory is expected to be applicable
for quenches in the ferromagnetic phase. It will turn out that it
actually holds in the whole ferromagnetic region not too close to the
critical point ($h=1$). In the paramagnetic phase one can start
with the $h\to\infty$ ground state to introduce an analogous
QP-concept involving individual spin flips instead of kinks
\cite{sachdev_young} but the same dispersion relation
(\ref{epsilon}) and velocity (\ref{vp}). We will mention the
necessary modifications below.

Immediately after the quench the time-dependent state of the system in
Eq.(\ref{Psi}), which for small $h$ and for small $t$ is given
by
\bea
|\varPsi_0(t)\rangle &\sim& \exp\left(-\imath t h\sum_l 
\sigma_l^x\right) |\varPsi_0\rangle \nonumber \\
&=&\prod_l\left[\cos(th)+\imath \sin(th) \sigma_l^x \right] 
|\varPsi_0\rangle
\;.
\label{Psi1}
\eea
(where we use the convention, $\sigma_i^x \leftrightarrow \sigma_i^z$,
as before). This indicates that by the action of the $\sigma_l^x$
operators initially single spins are flipped and thus pairs of kinks are
created at each lattice point, which then move ballistically with a
speed $v\sim h$. The maximum velocity is $v_{\rm max}\approx h$ for
small $h$.

In a translationally invariant system the creation probability of QP-s
is uniform and will be denoted by $f_p(h_0,h)$. For open boundary
conditions there will be corrections to a uniform creation probability
close the boundaries, which are negligible for sufficiently large
system sizes. In an equilibrated system that is thermalized at
temperature $T$, this would be $f_p^{\rm eq}(h_0,h)=e^{-\epsilon_p/T}$.
For zero temperature quantum relaxation $f_p(h_0,h)$ is the
probability with which the modes with momentum number $p$ are occupied
in the initial state $|\Psi_0\rangle$, i.e.\
\be
f_p(h_0,h)=\langle\Psi_0|\eta_p^+\eta_p|\Psi_0\rangle.
\label{fp}
\ee

\begin{figure}[t]
\begin{center}
\includegraphics[width=8cm]{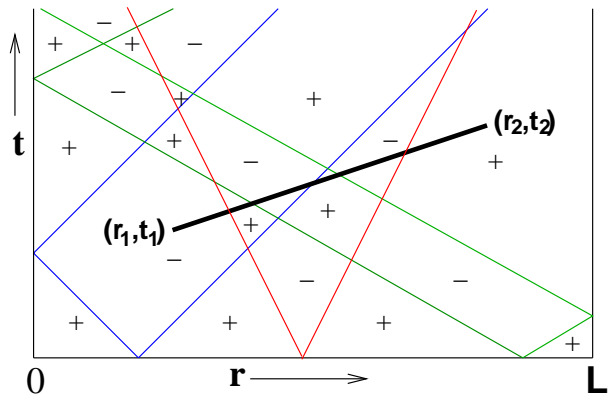}
\end{center}
\caption{
\label{fig-sketch-corr} 
(Color online) Typical semi-classical contribution to the correlation
function $C(r_1,t_1;r_2,t_2)$. Note the 6 trajectories of the
3 QP-pairs intersect the line $(r_1,t_1;r_2,t_2)$ five, i.e.\
an odd number, of times, which implies that $\sigma_{r_1}^x(t_1)$ and
$\sigma_{r_2}^x(t_2)$ have opposite orientation. Equivalently 
one can say that the trajectories of the red and the green QP pair 
intersect $(r_1,t_1;r_2,t_2)$ an even number of times (and thus do
not contribute) and the trajectory of the blue QP pair an odd number of times.
}
\end{figure}

In a finite system with open boundaries a QP with momentum $p$ moves
uniformly with velocity $v_p$ until it reaches one of the boundaries,
where it is reflected and moves with velocity $-v_p$ thereafter, and
so forth.  The trajectory of the kink is periodic in time, after a
time $2T_p$ with
\be
T_p=L/v_p
\ee
(including a reflection at the right and left boundary) it returns
to the starting point $x_0$ with the initial direction and velocity
$v_p$, see Fig. \ref{fig-sketch-corr}. Due to conservation of momenta
after a global quench QPs emerge pairwise at
random positions with velocities $+v_p$ and $-v_p$, as
indicated in Fig. \ref{fig-sketch-corr} for three QP pairs.

For a given QP pair created (at $t=0$) at position $x_0\in[0,L]$ let
$x_1(t)$ be the position of the initially right-moving QP (i.e\ with
initial velocity $v_p$) at time $t$ and $x_2(t)$ be the position of
the initially left-moving one (i.e. with initial velocity
$-v_p$). Define $t_a$ as the time when the left-moving particle
reaches the left wall the first time and $t_b$ as the time when the
right-moving particle reaches the right wall the first time
\bea
t_a & = & x_0/v_p\nonumber\\
t_b & = & (L-x_0)/v_p
\eea
Then for $t\le T_p$
\bea
x_1(t) & = & 
\left\{
\begin{array}{lcl}
x_0+v_p t   & \quad{\rm for}\; & t\le t_b\\
2L-x_0-v_pt & \quad{\rm for}\; & t_b<t\le T_p
\end{array}
\right.
\nonumber \\
x_2(t) & = & 
\left\{
\begin{array}{lcl}
x_0-v_p t & \quad{\rm for}\; & t\le t_a\\
v_pt-x_0  & \quad{\rm for}\; & t_a<t\le T_p
\end{array}
\right.
\label{x1x2}
\eea
At $t=T_p$ the two QPs meet at $x=L-x_0$. For $T_p<t<2T_p$ the
trajectories are defined accordingly (see Fig. \ref{fig-qp1}), and for
$t>2T_p$ one notes that $x_1$ and $x_2$ are $2T_P$-periodic.

Since QPs represent kinks or domain walls, $\sigma^x$ changes sign
each time a QP passes. Therefore the correlation function in
Eq.(\ref{C}) can be evaluated in terms of classical particles
moving according to (\ref{x1x2}) by using a similar reasoning
as in equilibrium \cite{sachdev_young}, the difference being 
that here 1) QP trajectories can intersect the line
$(r_1,t_1;r_2,t_2)$ several times, 2) QP trajectories come
always in pairs with a common off-spring at $t=0$, and 3) 
the occupation number of QPs is not thermal. 

If a QP trajectory intersects the line $(r_1,t_1;r_2,t_2)$ an odd
number of times, the spins at $(r_1,t_1)$ and $(r_2,t_2)$ have the
opposite orientations (i.e.\
$\sigma_{r_1}^x(t_1)=-\sigma_{r_2}^x(t_2)$), which contributes to the
decay of the correlation between $\sigma_{r_1}^x(t_1)$ and
$\sigma_{r_2}^x(t_2)$, see Fig.\
\ref{fig-sketch-corr}.
If the two trajectories pass an even number of times, the spins have
the same orientation, as if the trajectories did not pass the line
$(r_1,t_1;r_2,t_2)$ at all. Let $Q(r_1,t_1;r_2,t_2)$ be the
probability, that the QPs, which have started from the same site, have
passed the line $(r_1,t_1;r_2,t_2)$ a total odd number of times.  Then
the probability that for a given set of $n$ sites the kinks have
passed (for each site total odd times) this line is:
$Q^n(1-Q)^{L-n}$. Summing over all possibilities we have:
\bea
\dfrac{C(r_1,t_1;r_2,t_2)}{C_{\rm eq}(r_1,r_2)}
&=& \sum_{n=0}^L(-1)^nQ^n(1-Q)^{L-n}\dfrac{L!}{n!(L-n)!} \nonumber \\
&=&(1-2Q)^L \approx e^{-2Q(r_1,t_1;r_2,t_2)L}\;,
\label{CC}
\eea
where $C_{\rm eq}(r_1,r_2)$ is the equilibrium correlation function in
the initial state and in the last step we have used that the
probability, $Q(r_1,t_1;r_2,t_2)$ is small.  To calculate $Q$ one
should average over the QPs with momenta $p\in[-\pi,\pi]$, or
equivalently one can average over QP-pairs which is restricted to
$p\in[0,\pi]$. In this second method we have the expression:
\be
Q(r_1,t_1;r_2,t_2)=\frac{1}{2\pi}\int_0^\pi dp\;.
f_p(h_0,h)\cdot q_p(r_1,t_1;r_2,t_2)
\label{Q}
\ee
in terms of the occupation probability (see Eq.(\ref{fp})) and the
passing probability, $q_p(r_1,t_1;r_2,t_2)$.  This latter quantity
measures the probability that the two trajectories $x_1(t)$ and
$x_2(t)$ of any QP-pair with momentum $p$ intersect the line
$(r_1,t_1;r_2,t_2)$ together an odd number of times. The same
probability for a given QP-pair which is emitted at site $x_0\in[0,L]$
is denoted by $q_p(x_0|r_1,t_1;r_2,t_2)$. If we assume that the
generation of QPs at the quench is homogeneous in space then we
obtain:
\be
q_p(r_1,t_1;r_2,t_2)=\dfrac{1}{L}\int_0^L dx_0\,q_p(x_0|r_1,t_1;r_2,t_2)\;.
\label{qp}
\ee

In most cases of interest (see below) it is possible to provide an
analytical form for the function $q_p(r_1,t_1;r_2,t_2)$. If not, the
number of intersections can straightforwardly be determined
numerically and averaged over $x_0$, yielding $q_p(r_1,t_1;r_2,t_2)$
and thus $Q(r_1,t_1;r_2,t_2)$ in Eq.(\ref{Q}) and the correlation
function in Eq.(\ref{C}).

\section{Local magnetization}

\begin{figure}[t]
\begin{center}
\includegraphics[width=3cm]{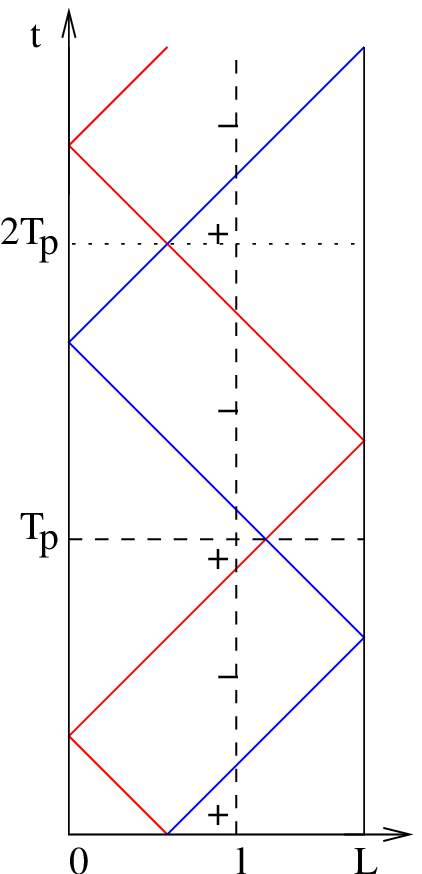}
\vspace{1.0cm}
\includegraphics[width=5cm]{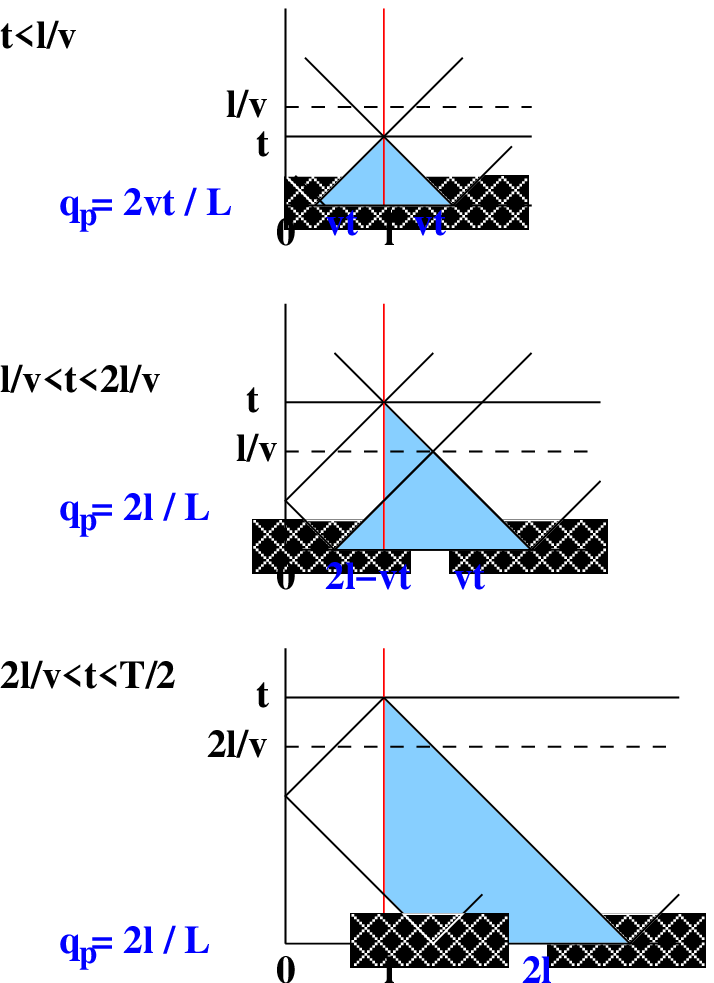}
\end{center}
\caption{
\label{fig-qp1} 
{\bf Left:} Typical semi-classical contribution to the time dependence
of the local magnetization $m_l(t)$.  Full lines are quasi-particles
or kinks moving with velocity $v_p$ through the chain. The $\pm$ signs
denote the sign of the spin at site $l$.  {\bf Right:} Sketch of the
trajectories of kink pairs that flip the spin at position $l$ exactly
once for times $t<T_p/2$. Kink pairs with initial position $x_0$ outside
the marked region either do not flip the spin at $l$ (since they do
not reach the position $l$ within time $t$) or they flip it twice.
$q_p$ is the fraction of the marked intervals on the $t=0$-axis. 
}
\end{figure}

The time-dependent local magnetization at a site $l$ (here we consider
$l \le L/2$) can be formally expressed to a correlation between a spin
that is fixed at time $t=0$ (to, say, $\sigma^x_l=+1$) and the same
spin at later times $t$, i.e.\ $m_l(t)=m_l^{\rm eq}\cdot
C|_{\sigma_l^x(t=0)=+}(l,0;l,t)$.  Then, with Eq.(\ref{C})
\be
m_l(t)=m_l^{\rm eq}\cdot e^{-2q(t,l)L}
\label{mt}
\ee
with $q(t,l)=Q|_{\sigma_l^x(t=0)=+1}(l,0;l,t)$, which is with Eq.(\ref{Q})
\be
q(t,l)=\frac{1}{2 \pi} \int_0^\pi dp \; f_p(h_0,h)\,q_p(t,l)
\label{qt}
\ee
where 
\be
q_p(t,l)=\dfrac{1}{L}\int_0^L dx_0\,q_p(x_0,t,l)
\ee
as in Eq.(\ref{qp}). To calculate $q_p$ one
concentrates first on times $t<T_p/2=L/2v_p$. Now $q_p$ is just
the fraction of possible initial positions from which kink pairs
can start with velocity $+v_p$ and $-v_p$ that flip the spin at
position $l$ exactly once. This region is marked in the sketch of
Fig. \ref{fig-qp1}. 
One sees that for $t<l/v_p$ one gets $q_p= 2vt/L$,
and for $l/v_p<t<T_p/2$ one gets $q_p=2l/L$, independent of
time.

For $T_p/2<t<T_p$ one observes that a kink pair that started (at $t=0$)
at position $x_0$ re-unites after a time $t=T_p$ at position
$L-x_0$. Since the origins of kink pairs are distributed uniformly over
the chain
the probability $q_p(t,l)$ is $T_p$-periodic
(n.b.: the kink trajectories themselves are only $2T_p$-periodic). 
Moreover $q_p(t,l)$ is symmetric with respect to time inversion
since it is symmetric under the QP velocity inversion, 
$q_p(-t,l)=q_p(t,l)$, therefore $q_p(T_p-t,l)=q_p(t,l)$. 
Defining the reflection times $t_1=l/v_p$ and $t_2=T_p-t_1$
one then has for the period $0\le t<T_p$ for $l<L/2$ 
\be
q_p(t,l)=\left\{
\begin{array}{lll}
2v_p t/L & \quad{\rm for}\quad & t\le t_1\\
2l/L     & \quad{\rm for}\quad & t_1\le t\le t_2\\
2-2v_p t/L & \quad{\rm for}\quad & t_2\le t<T_p
\end{array}
\right.
\label{qpt}
\ee
For $l>L/2$ one uses the symmetry $q_p(t,l)=q_p(t,L-l)$ and
for $t>T_p$ one makes use of the $T_p$-periodicity of $q_p(t,l)$: 
\be
q_p(t+nT_p,l)=q_p(t,l),\quad(n=1,2,\ldots).
\label{qptn}
\ee
Although $q_p(t,l)$ is $T_p$-periodic, $q(t,l)$ is not periodic,
since all QPs have different speed. Nevertheless the maximum speed
$v_{\rm max}=h+{\cal O}(h^2)$ determines the onset of 
magnetization reconstruction and therefore a quasi-periodicity of
$q(t,l)$ and concomitantly $m_l(t)$, whose (quasi)-period 
is then expected to be 
\be
T_{\rm period}=L/v_{\rm max}\approx L/h
\label{period}
\ee
With Eqs.(\ref{qt}), (\ref{qpt}) and (\ref{qptn})
one obtains $m_l(t)$ via numerical integration
(or summation over the discrete $p$-values for a lattice
of finite size $L$). 

For an actual calculation one needs to know the occupation probability
$f_p(h_0,h)$ in Eq.(\ref{fp}), which can be calculated numerically in a
straightforward manner using the free fermion technique. Since for
large system sizes the occupation probability is not expected to
depend strongly on the boundary condition we will use in the
following the expression for $f_p(h_0,h)$ for periodic boundary conditions, 
which can be given in analytical form as shown in Appendix A,
see Eq.(\ref{fp_appr}). In Fig.\ \ref{fig-mag-3d} the prediction 
of the semi-classical computation is shown. One observes the 
predicted quasi-periodicity for finite lattices and the
expected exponential decays in $l$ and $t$ as discussed below.

\begin{figure}[t]
\begin{center}
\includegraphics[width=8cm]{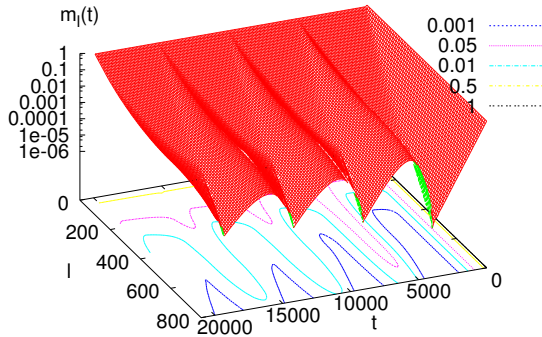}
\end{center}
\caption{
\label{fig-mag-3d}
Semi-classical prediction for the local magnetization $m_l(t)$
quench. Here $L=1024$, $h_0=0$, $h=0.2$. The (quasi)-periodicity
(\ref{period}) is $T_{\rm period}=L/h=5120$.}
\end{figure}

\begin{figure}[t]
\begin{center}
\includegraphics[width=\columnwidth]{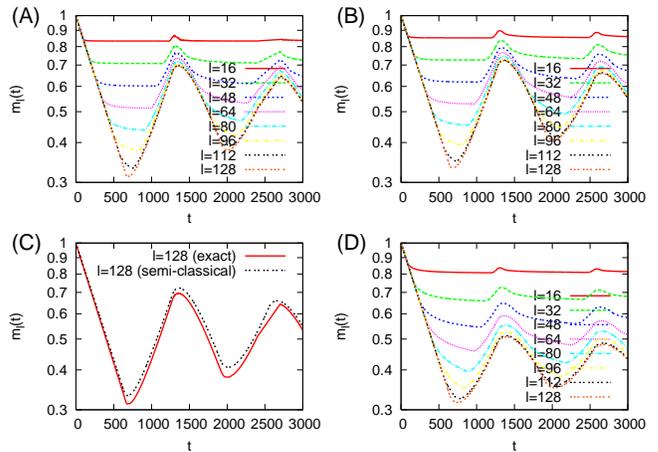}
\end{center}
\caption{
\label{fig-qp2} 
Relaxation of the local magnetization, $\log m_l(t)$, at different
positions in a $L=256$ chain with free ends after a quench with
parameters $h_0=0.0$, $h=0.2$ and $L=256$.  {\bf A} Exact (free
fermion calculation). {\bf B} Semi-classical prediction (\ref{mt})
with the passing probability (\ref{qpt}) and the occupation probability 
(\ref{fp_appr}). {\bf C} Comparison between exact and QP calculation for $m_l(t)$ for
$L=256$, $l=128$ for a quench from $h_0=0$ to $h=0.1$.
{\bf D} Semi-classical prediction using a thermal
occupation number probability in Eq.(\ref{fp-therm}) with an effective
temperature, $T_{\rm eff}$, see the text.}
\end{figure}

In Fig.\ \ref{fig-qp2} we compare the semi-classical prediction with
the exact results obtained with the free fermion technique and find
that the agreement is remarkably good.  We observe that small
deviations occur in the bulk ($l\sim L/2$) for $t>T_{\rm period}/2$,
which is when the first QP reflections are involved in the dynamical
evolution of $m_l(t)$. For sites close to the boundary, i.e.\ small
$l$ we observe small deviations already in the plateau region (c.f.\
also the surface-to-bulk correlations discussed further below). Here a
spatially inhomogeneous QP creation probabilities would have the most
significant effect, which is negligible for bulk spins.

When the system after the quench would be thermalized 
at some effective temperature $T_{\rm eff}(h_0,h)$ \cite{rossini}
this would imply that the occupation probability is 
\be
f_p(h_0,h)=e^{-\epsilon_p(h)/T_{\rm eff}(h_0,h)}\;.
\label{fp-therm}
\ee
The effective temperature is determined from the condition, that
the relaxation time in equilibrium $\tau_T(h,T)$ (with transverse
field $h$ and temperature $T$) is the same as in quantum relaxation at
$T=0$ but after a quench from $h_0$ to $h$.  In the limit $T \ll
\Delta(h)$, $\Delta(h)$ being the gap of the system we have\cite{sachdev_young}:
$\tau_T(h,T)\approx\dfrac{\pi}{2 T} e^{\Delta/T}$, which should be
compared with $\tau(h_0,h)$, which for small $h$ and $h_0$ is given in
Eq.(\ref{tau}). The result of the semi-classical calculation using
(\ref{fp-therm}) is also shown in Fig.\ \ref{fig-qp2}D and compared
with the exact data for a quench form $h_0=0$ to $h=0.2$. One sees
that, using the proper effective temperature the initial exponential
decay agrees perfectly, but soon as the first reflections 
are involved, large deviations occur. An effective
temperature can describe the initial relaxation well because
essentially it is a fit parameter for the initial exponential decay,
which stops after some ($l$-dependent) time in a finite system.

In the infinite system size limit $L\to\infty$ the time $t_2$ in
(\ref{qpt}) is infinite for all momenta $p$. Thus the
(quasi)-periodicity of $m_l(t)$ is lost and the functional form of
$m_l(t)$ as predicted by (\ref{qpt}), (\ref{qt}), and (\ref{mt}) is
\bea
m_l(t) = & m_l^{\rm eq} &
\exp\left(-t\cdot\frac2\pi\int_0^\pi dp\,v_p\,f_p(h_0,h)\,\theta(l-v_pt)\right)
\nonumber\\
& \cdot & 
\exp\left(-l\cdot\frac2\pi\int_0^\pi dp\,f_p(h_0,h)\,\theta(v_pt-l)\right)
\eea
which defines, in analogy to \cite{sachdev_young}
the quench specific length and time scales
\bea
\tau_{\rm mag}^{-1}(h_0,h) & = & 
\frac2\pi\int_0^\pi dp\,v_p\,f_p(h_0,h)\nonumber\\
\xi_{\rm mag}^{-1}(h_0,h) & = & 
\frac2\pi\int_0^\pi dp\,f_p(h_0,h)
\label{mag-tau}
\eea
In the small $h$ and $h_0$ limit these are calculated in Eqs.(\ref{tau}) and (\ref{xi_appr}), respectively.

In a finite system there is a quasi-periodicity and the magnetization
after the first relaxation period is reconstructed. Due to the
$p$-dependence of the velocity of the QPs in the reconstruction regime the
rate of exponential increase of the magnetization, $\tau'_{\rm mag}$, is increasing in time. Its maximal value is
reached at $t=T_{\rm period}$,
which is given by:
\bea
\dfrac{1}{\tau'_{\rm mag}(h_0,h)} & = & 
\frac2\pi\left[ \int_{\pi/6}^\pi - \int_{0}^{\pi/6}\right]dp\,v_p\,f_p(h_0,h) \nonumber\\
& \approx & h(h-h_0)^2 \dfrac{9\sqrt{3}-8}{12 \pi}
\label{mag-tau1}
\eea
where the second expression is valid in the small $h$ and $h_0$
limit. One can see that $\tau(h_0,h)<\tau'(h_0,h)$, thus the
reconstruction is slower than the relaxation. While QPs with large
energy and high velocity contribute to the
reconstruction, the other QPs with smaller energy and lower velocity
still reduce the magnetization. These processes with opposite effect
are responsible for the decay of the amplitude of the quasi-periodic
oscillations of the profile, see Figs.\ref{fig-mag-3d} and
\ref{fig-qp2}.

\begin{figure}[t]
\begin{center}
\includegraphics[width=\columnwidth]{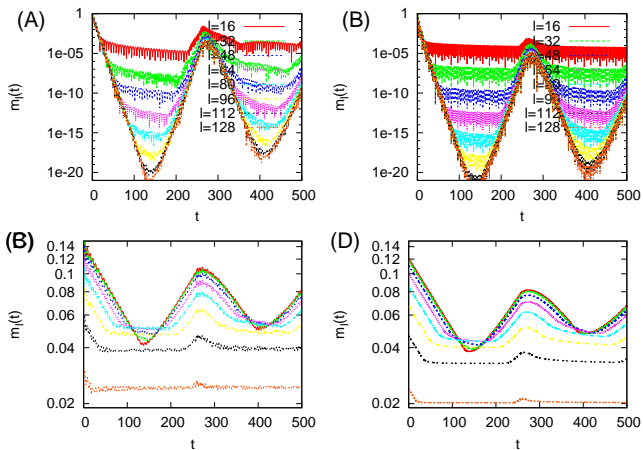}
\end{center}
\caption{
\label{fig-qp3}
Relaxation of the local magnetization, $m_l(t)$ after quenches 
into the disordered phase -- from the ordered phase 
($h_0=0.5$, $h=1.5$) {\bf (A)} exact, {\bf (B)} semi-classical, 
and from the disordered phase ($h_0=1.5$, $h=2.0$) 
{\bf (C)} exact, {\bf (D)} semi-classical. The legend of 
A and B holds also for C and D, the system size is $L=256$.
}
\end{figure}

After quenches into the disordered phase ($h>1$) 
the relaxation (and recurrent) dynamics of the 
longitudinal magnetization is superposed by oscillations
from the ground state correlations \cite{sachdev_young} and
one has to replace $m_l^{\rm eq}$ in (\ref{mt}) by
\be
m_l^{\rm eq} \to m_l^{\rm eq}\cdot K(t\Delta)\;,
\ee
where $K(x)$ is the modified Bessel function. The results for the
corresponding QP calculation and comparison with the exact data are
shown in Fig. \ref{fig-qp3}.  One observes again that the relaxation
and recurrent dynamics is well described by the semi-classical
picture also for quenches into the paramagnetic (disordered)
phase. The superposed oscillations have a slightly larger amplitude
and frequency.  Note also that for quenches from the paramagnetic
phase (Fig. \ref{fig-qp3}) C, D) the equilibrium profiles $m_l^{\rm
eq})$ shifts the curves for $m_l(t)$ downwards for increasing $l$,
since in the paramagnetic phase the surface magnetization is
larger than the bulk magnetization in a finite chain
(both vanishing only in the infinite system size limit).

\section{Correlation functions}

As mentioned before it is possible to perform the semi-classical
calculation for the two-spin correlations $C(r_1,t_1;r_2,t_2)$ for any
pair of sites $r_1$, $r_2$ and any pair of times $t_1$, $t_2$
with the formulas (\ref{C}), (\ref{Q}), and (\ref{qp}). Here we
want to focus on the time dependence of equal time correlations between spins
separated by a distance $r$ and arranged symmetrically within the bulk,
i.e.\ we consider
\be
C_t(r)=C(L/2-r/2,t;L/2+r/2,t)\;,
\ee
for quenches within the ordered phase ($h<1$),
which is, within the semi-classical theory given by:
\be
C_t(r)= C_{\rm eq}(r) \cdot \exp\left (- \frac{\textstyle L}{\textstyle 2\pi}
\int_0^\pi dp\; f_p(h_0,h)\cdot q_p^c(t,r)\right)
\label{corr}
\ee

\begin{figure}[t]
\begin{center}
\includegraphics[width=8cm]{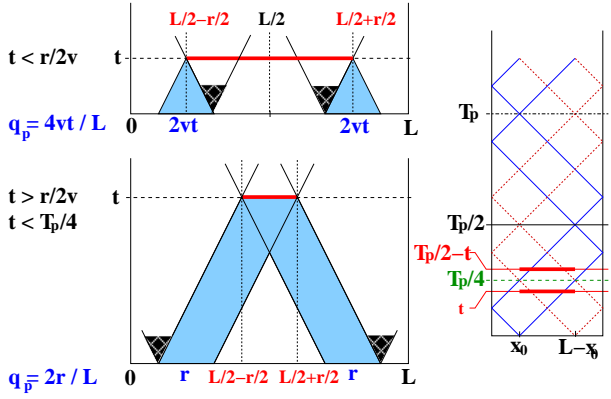}
\end{center}
\caption{
\label{fig-sketch-corr2}
(Color onlne) 
Semi-classical contributions to the equal time
correlation function $C_t(r)=C(L/2-r/2,t;L/2+r/2,t)$.
{\bf Left:} Sketch of the 
trajectories of kink pairs that reverse the orientation of 
the spins at $r_1$ and $r_2$ for times $t<T_p/4$.
Kink pairs with initial position $x_0$ outside
the marked region either do not intersect the 
line $(r_1=L/2-r/2,t;r_2=L/2+r/2,t)$ (red) (since they do 
not reach the red line within the time $t$) or they flip it twice. 
$q^c_p$ is the fraction of the marked intervals on the $t=0$-axis.
{\bf Right:} Sketch of the additional symmetry of $q^c_p(t)$
that reduce its periodicity from $T_p$ to $T_p/2$: For each 
QP pair created at position $x_0$ intersection the red line
at time $t<T_p/4$ there is a QP pair created at position 
$L-x_0$ that intersects the red line at time $T_p/2-t$.
Hence $q^c_p(t)=q^c_p(T_p/2-t)$ for $t<T_p/2$. At $t=T_p/2$
the QP pair created at $x_0$ meets again at $L-x_0$ and
the one created at $L-x_0$ meets again at $x_0$, which 
implies after averaging over initial position that 
$q^c_p(t+T_p/2)=q^c_p(t)$.
}
\end{figure}

\begin{figure}[t]
\begin{center}
\includegraphics[width=\columnwidth]{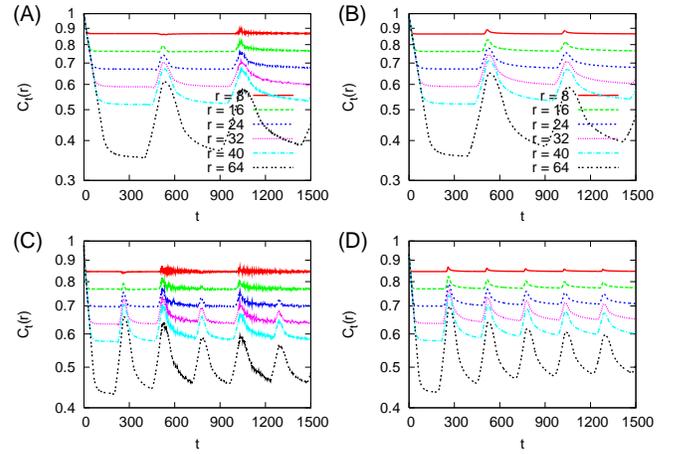}
\end{center}
\caption{
\label{fig-qp4}
Equal time correlation function $C_t(r)$ for fixed $r$
as a function of time $t$ after the quench:
Comparison between the exact result (left) and the 
semi-classical prediction (right). $L=256$,
$h_0=0.$, $h=1.5$ {\bf (A)} exact, {\bf (B)} semi-classical;
$h_0=0.3$, $h=0.5$ {\bf (C)} exact, {\bf (D)} semi-classical;
The legend of A and B holds also for C and D.}
\end{figure}

\begin{figure}[t]
\begin{center}
\includegraphics[width=\columnwidth]{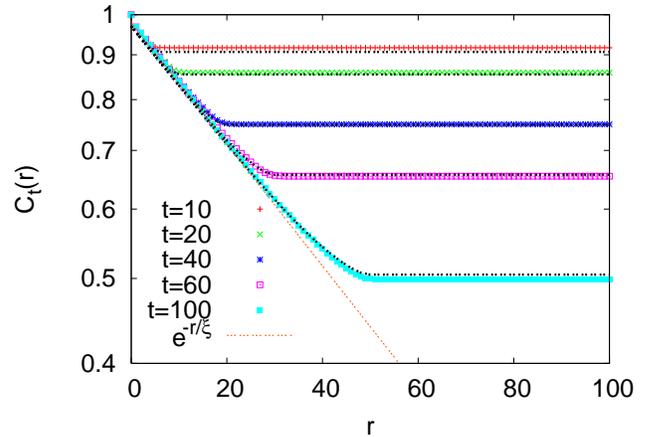}
\end{center}
\caption{
\label{fig-qp5}
Equal time correlation function $C_t(r)$ for fixed 
time $t$ after the quench as a function of distance $r$ 
Comparison between the exact result (points) and the 
QP calculation (black broken lines). 
$L=256$, $h_0=0$, and $h=0.25$.
}
\end{figure}

As sketched in Fig. \ref{fig-sketch-corr2} the function
$q^c_p(r,t)$ for $C_t(r)$ is $T_p/2$-periodic and for the
period $0\le t<T_p/2$ given by for $r<L/2$
\be
q^c_p(t,r)=\left\{
\begin{array}{lll}
4v_p t/L & \quad{\rm for}\quad & t\le t_1\\
2r/L     & \quad{\rm for}\quad & t_1\le t\le t_2\\
2-4v_p t/L & \quad{\rm for}\quad & t_2\le t<T_p/2
\end{array}
\right.
\label{qptc}
\ee
with $t_1=r/2v_p$ and $t_2=T_p/2 - t_1$. (For $r>L/2$ one should
replace in the above formulas $r$ to $L-r$.) Note that the relevant
times occurring in this expression are all multiplied with a factor
$1/2$ as compared to those determining $q_p$ for the local
magnetization (\ref{qpt}). In particular $q_p(t,r)$ is
$T_p/2$-periodic (in contrast to the $T_p$ periodicity of $q_p$ for
the local magnetization): $q_p(t+nT_p/2,r)=q_p(t,r)$ for
$n=1,2,3,\ldots$. As a result the (quasi)-period of $C_t(r)$ for fixed
$r$ is one half of the (quasi)-period of the local magnetization 
$m_l(t)$
\be
T_{\rm period}^C=L/2v_{\rm max}\approx L/2h\;.
\label{tpc}
\ee
With (\ref{qptc}) and $f_p(h_0,h)$ from appendix A the semi-classical
calculation can be performed, results and the comparison with exact
data are shown in Fig. \ref{fig-qp4} and \ref{fig-qp5}.

\begin{figure}[t]
\begin{center}
\includegraphics[width=\columnwidth]{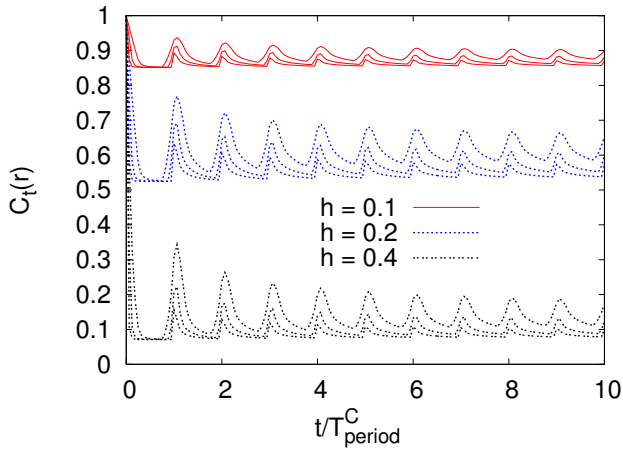}
\end{center}
\caption{
\label{fig-qp6}
QP prediction for the equal time correlation function $C_t(r)$
plotted against time after the quench scaled with the
(quasi)-period $T_{\rm period}^C=L/2h$ for different fields $h$
($h_0=0$). The three curves for each field $h$ correspond to 
different system sizes: $L=256$, $512$, and $1024$ 
(from the top curve to the bottom curve).}
\end{figure}

In \ref{fig-qp6} we show the semi-classical prediction for $C_t(r)$
for larger system sizes and long times, scaled by the 
(quasi)-period $T_{\rm period}^C$, (\ref{tpc}), which demonstrates 
the persistence of the recurrence for very long times
in finite systems. Note that the recurrence amplitude decreases 
with increasing system size and vanishes completely for $L\to\infty$.

In the infinite system size limit $L\to\infty$ the time $t_2$ in
(\ref{qpt}) is infinite for all momenta $p$. Thus the
(quasi)-periodicity of $C_t(r)$ is lost and the functional form of
$C_t(r)$ as predicted by (\ref{qptc}) and (\ref{corr}), is
\bea
C_t(r) = & C_{\rm eq}(r) &
\exp\left(-t\cdot\frac4\pi\int_0^\pi dp\,v_p\,f_p\,\theta(l-v_pt)\right)
\nonumber\\
& \cdot & 
\exp\left(-r\cdot\frac2\pi\int_0^\pi dp\,f_p\,\theta(v_pt-l)\right)\;,
\label{crt-inf}
\eea
with $f_p=f_p(h_0,h)$. This agrees to first order in $f_p$ to the
prediction of \cite{Calabrese2}, where $f_p$ is replaced by
$-1/2\;\log(1-2f_p)=f_p+{\cal O}(f_p^2)$ (see appendix
A). Eq. (\ref{crt-inf}) defines the quench specific length and time
scales
\bea
\tau_c^{-1}(h_0,h) & = & 
\frac4\pi\int_0^\pi dp\,v_p\;f_p(h_0,h)\nonumber\\
\xi_c^{-1}(h_0,h) & = & 
\frac2\pi\int_0^\pi dp\,f_p(h_0,h)
\label{corr-c}
\eea
Note that $\tau_c=\tau_{\rm mag}/2$ and $\xi_c=\xi_{\rm mag}$.
For a small $h_0$ and $h$ this yields to leading order
\be
\xi^{-1}(h_0,h)=\dfrac{(h-h_0)^2}{2\pi}\int_0^{\pi} {\rm d}k \sin^2 k=\dfrac{(h-h_0)^2}{4}
\label{xi_appr}
\ee

\subsection{Surface-to-bulk correlation}

The surface-to-bulk correlation function 
$C_t^{\rm surf}(r)=C(0,t;r,t)$ is within
semi-classical theory given by
\be
C_t^{\rm surf}(r)= C_{\rm eq}^{\rm surf}(r)
 \cdot \exp\left (-\frac{\textstyle L}{\textstyle\pi}
\int_0^\pi dp\,f_p\cdot q^{\rm surf}_p(t,r)\right)
\ee
with $f_p=f_p(h_0,h)$.  Similar considerations that lead to
(\ref{qpt}) and (\ref{qptc}) yield an analytical expression for
$q_p^{\rm surf}(t,r)$, which is equivalent to $q_p(t,l)$ for the local
magnetization in (\ref{qpt}), however with $l=r$:
\be
q_p^{\rm surf}(t,r)=q_p(t,l=r)\;.
\ee
Consequently
\be
C_t^{\rm surf}(r)=C_{\rm eq}^{\rm surf}(r)
\frac{m_{l=r}(t)}{m_{l=r}^{\rm eq}}\;.
\ee
which implies that the surface-bulk correlation is dominated by the
relaxation of the magnetization at the bulk site and that it is
$T_p$-periodic (in contrast to $C_t(r)$, which is $T_p/2$-periodic.

\begin{figure}[t]
\begin{center}
\includegraphics[width=\columnwidth]{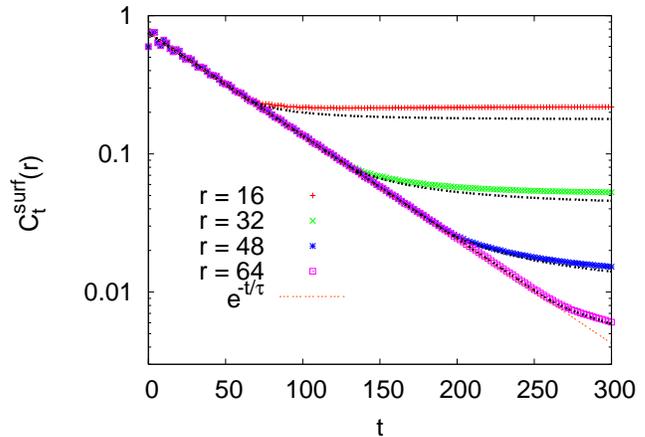}
\end{center}
\caption{
\label{fig-corr-surf}
Surface-to-bulk correlation function $C_t^{\rm surf}(r)$ for fixed 
time $r$ as a function of the time after the quench.
Comparison between the exact result (points) and the 
QP calculation (black broken lines). 
$L=256$, $h_0=0.75$, and $h=0.25$.
}
\end{figure}

In Fig.\ \ref{fig-corr-surf} we show a comparison of this
semi-classical result with the exact data. We observe that small
deviations occur in the plateau region for small distances $r$ where
the bulk-correlations still agree very well with the semi-classical
prediction. Since for small $r$ both sites in the 
surface-to-bulk correlation function are close to the boundary 
a spatially inhomogeneous QP creation probabilities would have 
the most significant effects here.

\subsection{Autocorrelations}

The autocorrelation function 
\be
G_l(t)=C(l,0;l,t)
\ee
is (up to an extra factor $m_l^{\rm eq}$) identical to
the time-dependent local magnetization $m_l(t)$:
\be
G_l(t)=G_{\rm eq}\frac{m_l(t)}{m_l^{\rm eq}}\;.
\ee
For $l=L/2$ (bulk autocorrelation) in the limit $L\to\infty$
the QP prediction is
\bea
G_{L/2}(t)& \propto & \exp\left(-\frac{2}{\pi}\int_0^\pi dp\;
f_p(h_0,h)\cdot v_p t\right)\nonumber\\
& = & e^{-t/\tau_{\rm auto}}
\eea
with the relaxation time $\tau_{\rm auto}=\tau_{\rm mag}$,
eq.(\ref{mag-tau}), which corresponds to the leading order of the
result from Calabrese \textit{et al} \cite{Calabrese2}(see appendix
A), in which $f_p$ is again replaced by $-1/2\;\log(1-2f_p)=f_p+{\cal
O}(f_p^2)$.  For a small $h_0$ and $h$ (\ref{mag-tau}) yields to
leading order
\be
\tau^{-1}=\dfrac{h(h-h_0)^2}{2\pi}\int_0^{\pi} {\rm d}k \sin^3 k=h(h-h_0)^2\dfrac{2}{3 \pi}\;.
\label{tau}
\ee
This has already been found numerically in\cite{igloi_rieger2}.
\vfill
\eject

\section{Periodic boundary conditions}

In a chain with periodic boundary conditions instead of the open
boundaries that we considered so far, one has to replace 
the QP trajectories (\ref{x1x2}) by the appropriate 
expressions:
\bea
x_1(t)=(x_0+v_pt)\;{\rm mod}\;L\nonumber\\
x_2(t)=(x_0-v_pt)\;{\rm mod}\;L
\eea
where the modulo operation is defined in the obvious manner:
Shift the real number $x_i$ by multiples of $L$ such that it
lays in the interval $[0,L]$. With this the evaluation of 
the local magnetization and correlation functions is straight-forward.

The chain with periodic boundary conditions is translationally
invariant, therefore the equal time correlation $C_t^{\rm
p.b.c.}(r)=C^{\rm p.b.c.}(r_1,t_1;r_1+r,t_2)$ is independent of
$r_1$. One sees immediately that the expression for $q_p^{\rm
p.b.c.}(t,r)$ is identical to $q_p^{c}(t,r)$ in
Eq. (\ref{qptc}), and therefore $C_t^{\rm p.b.c}(r)$ is identical,
up to prefactors from the ground state or equilibrium 
correlation function, to $C^{\rm open}(L/2-r/2,t;L/2+r/2,t)$:
\be
C_t^{\rm p.b.c.}(r)
=\frac{C_{\rm eq}^{\rm p.b.c.}(r)}{C_{\rm eq}^{\rm open}(r)}
\cdot C_t^{\rm open}(r)
\ee
It should be noted that this relation only holds for
the symmetric correlation function
$C_t^{\rm open}(r)=C^{\rm open}(L/2-r/2,t;L/2+r/2,t)$.

The local magnetization $m_l(t)$ is independent of the site $l$
in a system with periodic boundary conditions, as is the
QP passing probability $q_p^{\rm p.b.c.}(t,l)=q_p^{\rm p.b.c.}(t)$.
We find
\be
q_p^{\rm p.b.c.}(t)=\left\{
\begin{array}{lll}
2v_p t/L & \quad{\rm for}\quad & t\le T_p/2\\
2-2v_p t/L & \quad{\rm for}\quad & T_p/2<t\le T_p
\end{array}
\right.
\label{qpbct}
\ee
For $t>T_p$ one uses the $T_p$-periodicity 
$q_p^{\rm p.b.c.}(t+nT_p)=q_p^{\rm p.b.c.}(t)$, $(n=1,2,\ldots)$.
With eq. (\ref{mt}) and (\ref{qt}) the local magnetization
is then given by
\bea
& & \frac{m^{\rm p.b.c.}(t)}{m_{\rm eq}^{\rm p.b.c.}} =
\exp \left( -\frac2\pi\int_0^\pi dp\;f_p\,v_pt\,\theta(\sin(2\pi t/T_p))
\right.\nonumber\\
& & \left.\qquad 
+ \frac2\pi\int_0^\pi dp\;f_p\,(L-v_pt)\,\theta(-\sin(2\pi t/T_p))
\right)
\label{mag-pbc}
\eea
In the infinite system size limit this yields 
\be
m^{\rm p.b.c.}(t)\propto
\exp \left( -t\frac2\pi\int_0^\pi dp\;f_p\,v_p\right)
\propto e^{-t/\tau_{\rm mag}}\;,
\ee
which agrees to first order in $f_p$ with the prediction 
of \cite{Calabrese2}. As for open boundary conditions
the autoscorrelation function $G^{\rm p.b.c.}(t)$ is
given, up to prefactors, by the same expression as the
local magnetization (\ref{mag-pbc}).

\section{Discussion}
\label{sec:disc}

We have formulated a semi-classical theory for the non-equilibrium
quantum relaxation of the transverse Ising chain after a global quench
via an instantaneous change of the transverse field.  It is applicable
to systems of finite and infinite length and describes properly the
relaxation dynamics as well as the recurrence / reconstruction
properties of dynamical correlations in finite systems. For infinite
systems our theory agrees to lowest order with a recent prediction by
Calabrese et al\cite{Calabrese2}.  Our results indicate that the
global quantum quench induces a unique length-scale, $\xi$, and a
unique time-scale, $\tau$, in the system, both dependent upon the quench
parameters, $h_0$ and $h$.  These characteristic scales appear also in
half-infinite geometry and in finite systems, provided the length of
the system is larger than $\xi$. In a finite system this
semi-classical theory not only explains the recurrence and
reconstruction properties of the local
magnetization\cite{igloi_rieger2}, but describes the dynamical
behavior quantitatively.

The semi-classical theory can be used to define an effective
temperature for the quantum relaxation process. If we compare the
expressions obtained by Sachdev and Young\cite{sachdev_young} for the
correlation length and the relaxation time in equilibrium at
finite temperatures with our results for zero temperature
quantum quenches one obtains a node-dependent effective
temperature, $T_{\rm eff}(p)$, defined by the condition:
\be
f_p(h_0,h)=\exp\left(-\dfrac{\varepsilon_h(p)}{T_{\rm eff}(p)}\right)\;.
\label{fp_boltzmann}
\ee
This relation agrees to first order in $f_p$ with the prediction of
\cite{Calabrese2} (i.e.\ for small effective temperatures or small
differences $|h-h_0|$). In \cite{Calabrese2} the Boltzmann-factor on
the r.h.s. of Eq.(\ref{fp_boltzmann}) is replaced by the
Fermi-function with zero chemical potential, as shown in the Appendix in Eq.(\ref{fp_fermi}),
thus replacing classical kinks simply by free fermions. 

It is interesting to notice an analogous expression for the
time-evolution of the entanglement entropy, $S(t)$, measured after the quench between two semi-infinite
parts of the system, say ${\cal A}$ and ${\cal B}$. The analytical result by Fagotti and
Calabrese\cite{fagotti_calabrese} can be written into the form:
\be
S(t)=t\frac1\pi\int_0^\pi dp\,v_p\,s_p(h,h_0)\;,
\label{S_t}
\ee
with

\be
s_p(h,h_0)=-(1-f_p)\ln(1-f_p)-f_p\ln f_p\;,
\ee
being the entropy of the fermionic mode with occupation number $f_p(h_0,h)$.
In the semi-classical theory this expression can be interpreted as the result of ballistically moving QP pairs,
which are created say at ${\cal A}$ at $t=0$ and one of them is reaching ${\cal B}$ before the actual time, $t$.
Each of these QPs brings an entropic contribution as a free fermion.  
It would be interesting to see if the relations in Eqs.(\ref{fp_fermi}) and (\ref{S_t}) are valid
for another integrable quantum spin systems, too.

The semi-classical approach is accurate, if the occupation
probability, $f_p(h_0,h)$, is small, which is valid if the initial and
the finite states are close to each other and both are
ferromagnetic. As shown in\cite{igloi_rieger2} for $h,h_0<1$ the
magnetization profile $m_l(t)$ for any finite $l$ and $t$ is
non-negative. In the other domains of the quench ($h_0$ and/or $h$ is
larger than $1$) during relaxation $m_l(t)$ takes negative values,
too. This type of oscillating relaxation is described qualitatively
well by the semi-classical theory.  The amplitude of the oscillations
as well as the recurrence of the magnetization and the correlations
are correctly described, but there are differences in the actual value
of the frequencies. For quenches close to the critical point we expect
the concept of isolated QPs to become invalid or at least
quantitatively inaccurate due to the diverging correlation length
either in the initial and / or final state.

In finite system with open boundaries and for half-infinite systems we
find small deviations between the exact and the semi-classical results
either when sites close to the boundaries are involved or for times
$t>T_{\rm period}/2$, when QPs reflected at the boundaries contribute
to the magnetization or correlation reconstruction.  A possible source
for the deviations in the first case is the lack of translational
invariance in chains with open boundaries, which results in spatially
inhomogeneous creation probability of QP pairs, at least close to the
boundaries. The second kind of deviations could originate in
the dynamical processes during the reflection at the open 
boundaries, which might be more complicated than just momentum
inversion. Both effects are absent in systems with open boundaries, 
for which reason we expect our predictions to be accurate 
for all times in finite chains with periodic boundary conditions. 

Our semi-classical theory can be generalized to several
directions. This theory is also valid for transverse Ising chains
involving a sum over more ferromagnetic short-range interactions than
only nearest neighbors, as has been argued for the equilibrium
relaxation dynamics at finite temperatures by Sachdev and
Young\cite{sachdev_young}. The semi-classical theory should be
applicable to non-integrable models, too, for which one has to include
QP-collision and scattering processes. Here the quantum Boltzmann
equation seems to be a promising approach \cite{ziman}, as has been
demonstrated recently for a bosonic system in \cite{rosch}.

\section*{Appendix}
\appendix

Here we compare our semi-classical calculation with the predictions by
Calabrese \textit{et al}\cite{Calabrese2}.  First we recapitulate the exact solution of
the free fermion representation the Hamiltonian in
Eq.(\ref{hamilton_free}) for periodic boundary conditions.  In this
case there are pairs of fermions with quasi-momenta $p$ and $-p$, and
in the ground state sector these are: $p=\dfrac{\pi}{L},
\dfrac{3\pi}{L}, \dfrac{5\pi}{L}, \dots$, $0<p<\pi$.  Here we define
the functions:
\bea
u_h(p)&=&\sqrt{\dfrac{\varepsilon_h(p)+h-\cos p}{2\varepsilon_h(p)}} \nonumber \\
v_h(p)&=&\sqrt{\dfrac{\varepsilon_h(p)-(h-\cos p)}{2\varepsilon_h(p)}}
\eea
and
\bea
U_p=u_{h_0}(p)u_h(p)+v_{h_0}(p)v_h(p)\\
V_p=u_{h_0}(p)v_h(p)-v_{h_0}(p)u_h(p)
\eea
in terms of which the ground state for $t<0$ ($|\Psi_0 \rangle$) is expressed with the
ground state at $t>0$ ($| 0 \rangle$) as:
\be
|\Psi_0 \rangle= \prod_p \left[ U_p + i V_p \eta_p^{\dag}\eta_{-p}^{\dag}\right]| 0 \rangle 
\ee
Then the density of quasi-particle excitations is given by the nonequilibrium occupation number:
\be
f_p=\langle \Psi_0|\eta_p^{\dag}\eta_p |\Psi_0 \rangle=|V_p|^2
\ee
This can be expressed as:
\be
f_p=\dfrac{1}{2}\left[1-\cos \Delta_p \right]
\ee
where $\Delta_p$ is the difference between the Bogoliubov angles diagonalizing ${\cal H}(h)$ and ${\cal H}(h_0)$, respectively: 
\be
\cos \Delta_p=\dfrac{h_0h-(h_0+h)\cos p +1}{\varepsilon_{h_0}(p)\varepsilon_h(p)}
\ee
If the difference between $h_0$ and $h$ is small we obtain in leading order for the occupation number:
\be
f_p=\dfrac{1}{4}(h-h_0)^2 \sin^2 p
\label{fp_appr}
\ee
The results by Calabrese \textit{et al}\cite{Calabrese2} can be formally obtained from our
semi-classical expressions, if an effective occupation number is used. For example
in Eq.(\ref{xi_appr}) for the correlation length and in Eq.(\ref{tau}) for the relaxation time one should
simply replace:
\be
f_p \to -\dfrac{1}{2}\ln|\cos \Delta_p|
\ee
The semi-classical results then represent the leading term of the exact expressions.

According to Calabrese \textit{et al}\cite{Calabrese2} there is an effective
thermal (Gibbs) distribution or generalized Gibbs ensemble (GGE),
which is obtained in integrable models by maximizing the entropy,
while keeping the energy and other conservation laws fixed.  This
leads to an effective, node-dependent temperature: $T_{\rm eff}(p)$,
which is given by:
\be
\cos \Delta_p=\tanh\dfrac{\epsilon_h(p)}{2T_{\rm eff}(p)}\;,
\ee
or expressed with $f_p$ we have:
\be
f_p=\dfrac{1}{\exp\left(\dfrac{\varepsilon_h(p)}{T_{\rm eff}(p)}\right) +1}.
\label{fp_fermi}
\ee
At the r.h.s. we have the Fermi distribution function with zero chemical potential, thus the GGE
condition is expressed in the form, that the nonequilibrium occupation
number of the given mode is equal to its thermal occupation at the
effective temperature: $T_{\rm eff}(p)$.

\begin{acknowledgments}
This work has been supported by the Deutsche Forschungsgemeinschaft
(DFG) and by the Hungarian National Research Fund under grant No OTKA
K62588, K75324 and K77629 and by a German-Hungarian exchange program
(DFG-MTA).
\end{acknowledgments}


\begin{thebibliography}{99}

\bibitem{barouch_mccoy}
E. Barouch and B. McCoy, Phys. Rev. A \textbf{2}, 1075 (1970);
Phys. Rev. A \textbf{3}, 786 (1971);
Phys. Rev. A \textbf{3}, 2137 (1971).

\bibitem{igloi_rieger}
F. Igl\'oi and H. Rieger,
Phys. Rev. Lett. \textbf{85}, 3233 (2000).

\bibitem{sengupta}
K. Sengupta, S. Powell and S. Sachdev,
Phys. Rev. A \textbf{69}, 053616 (2004).

\bibitem{exp}
M. Greiner, O. Mandel, T. W. H\"ansch and I. Bloch,
Nature \textbf{419}, 51 (2002);
L. E. Sadler, J. M. Higbie, S. R. Leslie, M. Vengalattore,
 and D. M. Stamper-Kurn, Nature \textbf{443} 312 (2006);
A. Lamacraf, Phys. Rev. Lett. \textbf{98}, 160404 (2006);
B. Paredes \textit{et al}.
Nature \textbf{429}, 277 (2004);
T. Kinoshita, T. Wenger and D. S. Weiss,
Science \textbf{305}, 1125 (2004);
T. Kinoshita, T. Wenger and D. S. Weiss,
Nature \textbf{440}, 900 (2006).

\bibitem{rigol}
M. Rigol, V. Dunjko, V. Yurovsky and M. Olshanii,
Phys. Rev. Lett. \textbf{98}, 50405 (2007).

\bibitem{kollath}
C. Kollath, A. L\"auchli, and E. Altman,
Phys. Rev. Lett. \textbf{98}, 180601 (2007).

\bibitem{roux}
G. Roux, Phys. Rev. A \textbf{79}, 021608(R) (2009).

\bibitem{gritsev}
V. Gritsev, E. Demler, M. Lukin, and A. Polkovnikov,
Phys. Rev. Lett. \textbf{99}, 200404 (2007).

\bibitem{cazalilla}
M. A. Cazalilla, Phys. Rev. Lett. \textbf{97}, 156403 (2006).

\bibitem{manmana}
S. R. Manmana, S. Wessel, R. M. Noack and A. Muramatsu,
Phys. Rev. Lett. \textbf{98}, 210405 (2007).

\bibitem{calabrese_cardy}
P. Calabrese and J. Cardy,
Phys. Rev. Lett. {\bf 96}, 136801 (2006);
J. Stat. Mech. P06008 (2007).

\bibitem{sotiriadis_cardy}
S. Sotiriadis and J. Cardy, J. Stat. Mech. P11003 (2008);
Phys. Rev. B \textbf{81}, 134305 (2010).

\bibitem{gambassi}
A. Gambassi and P. Calabrese,
arXiv:1012.5294 (2010).

\bibitem{rossini}
D. Rossini, A. Silva, G. Mussardo, and G. E. Santoro,
Phys. Rev. Lett. \textbf{102}, 127204 (2009);
D. Rossini, S. Suzuki, G. Mussardo, G. E. Santoro and A. Silva
Phys. Rev. B \textbf{82}, 144302 (2010).

\bibitem{igloi_rieger2}
F. Igl\'oi and H. Rieger,
Phys. Rev. Lett. {\bf 106}, 035701 (2011)

\bibitem{Calabrese2}
P. Calabrese, F. H. L. Essler, M. Fagotti,
Phys. Rev. Lett. {\bf 106}, 227203 (2011).

\bibitem{sachdev_young}
A. Sachdev and A. P. Young, Phys. Rev. Lett. 78, 2220 (1997).

\bibitem{calabrese_cardy2}
P. Calabrese and J. Cardy,
J. Stat. Mech. P04010 (2005).

\bibitem{fagotti_calabrese}
M. Fagotti and P. Calabrese, Phys. Rev. A \textbf{78}, 010306(R) (2008).

\bibitem{gradient}
V. Eisler, F. Igl\'oi and I. Peschel, J. Stat. Mech. P02011 (2009).

\bibitem{uma}
U. Divakaran, F. Igl\'oi, H. Rieger, arXiv:1105.5317 (2011).

\bibitem{pfeuty} P. Pfeuty, Ann. Phys. \textbf{57}, 79 (1970).

\bibitem{lieb} E. Lieb, T. Schultz, and D. Mattis, Ann. Phys. \textbf{16}, 407 (1961).

\bibitem{yang} C. N. Yang, Phys. Rev. 85, 808 (1952).

\bibitem{ziman}
J. M. Ziman, Electrons and Phonons (Oxford University
Press, New York, 1960).

\bibitem{rosch}
U. Schneider et al., arXiv:1005.3545 (2010).



\end{thebibliography}
\end{document}